\documentclass[11pt,twoside]{article}
\usepackage[utf8]{inputenc}
\usepackage{times,fancyhdr}
\usepackage[dvips]{graphicx}
\usepackage{latexsym}
\usepackage[affil-it]{authblk}
\usepackage{amsmath}
\usepackage{amssymb}
\usepackage{mathtools}
\usepackage[title, titletoc]{appendix}

\usepackage[dvipsnames]{xcolor}
\usepackage[colorlinks,allcolors=Blue]{hyperref}

\usepackage[backend=biber,style=phys,citestyle=authoryear-comp,biblabel=brackets,pageranges=false,eprint=true,natbib=true]{biblatex}
\DeclareSourcemap{
 \maps[datatype=bibtex,overwrite=true]{
  \map{
    \step[fieldsource=Collaboration, final=true]
    \step[fieldset=usera, origfieldval, final=true]
  }
 }
}

\renewbibmacro*{author}{%
  \printnames{author}
  \iffieldundef{usera}{
  }{
    (\printfield{usera} Collaboration)
  }
}

\DeclareFieldFormat{eprint:arxiv}{%
    arXiv\addcolon
    \ifhyperref
    {\href{https://arxiv.org/abs/#1}{%
        #1%
    }}
    {%
        #1%
    }}

\AtEveryBibitem{\clearfield{eid}}

\usepackage[top=3.93cm, bottom=3.93cm, left=4cm, right=4cm]{geometry}

\begin{document}

\title{Comment on ``Observational Evidence for Cosmological Coupling of Black Holes and its Implications for an Astrophysical Source of Dark Energy''}
\author{Tobias Mistele\thanks{\href{mailto:tobias.mistele@case.edu}{tobias.mistele@case.edu}}}
\affil{\small Department of Astronomy, Case Western Reserve University\\
10900 Euclid Avenue,
44106 Cleveland, Ohio, USA
}
\date{}
\maketitle

\begin{abstract}
It was recently claimed that black holes can explain the accelerated expansion of the universe.
Here I point out that this claim is based on a confusion about  the principle of least action, undermining the link between black holes and dark energy.
\end{abstract}

\section{The original argument}

In recent work, \citet{Farrah2023, Farrah2023b} find observational evidence that black hole masses grow with the cosmic scale factor $a$ as $a^{3}$ (but see \citet{Rodriguez2023}).
This finding is then taken to show that black holes are responsible for the accelerated expansion of the universe.
That is, the claim is that black holes are the dark energy.

The link between black hole growth and cosmic acceleration is based on earlier theoretical work of \citet{Croker2019}.
They give an argument for why black holes can lead to cosmic acceleration and argue that a testable implication of their argument is that black hole masses grow as $a^3$.
This argument goes as follows.

First, \citet{Croker2019} derive a Friedmann-like equation for the effective cosmological scale factor $a$ in the presence of matter inhomogeneities,
\begin{align}
 \label{eq:friedmann}
 \partial_\eta^2 a(\eta) = \frac{4 \pi G}{3} a^3(\eta) \Bigl< \rho(\eta, \vec{x}) - \sum_{i=1}^3 P_i(\eta, \vec{x}) \Bigr>_V \,,
\end{align}
where $\eta$ is conformal time and $\rho$ and $P$ are the -- inhomogeneous -- density and pressure of matter.
The averaging denoted by $\langle \dots \rangle$ is defined as
\begin{align}
  \Bigl< \rho(\eta, \vec{x}) - \sum_{i=1}^3 P_i(\eta, \vec{x}) \Bigr>_V = \frac{1}{V} \int_V d^3\vec{x} \left(\rho(\eta, \vec{x}) - \sum_{i=1}^3 P_i(\eta, \vec{x}) \right) \,.
\end{align}
The spatial volume $V$ is taken to be large but finite in \citet{Croker2019}.
The Friedmann-like equation Eq.~\eqref{eq:friedmann} implies that the cosmologically relevant density and pressure are simple spatial averages,
\begin{align}
  \label{eq:avgs}
   \rho_{\mathrm{eff}}(\eta) = \Bigl<\rho(\eta, \vec{x})\Bigr>_V \,, \quad
   3 P_{\mathrm{eff}}(\eta) = \Bigl<\sum_{i=1}^3 P_i(\eta, \vec{x}) \Bigr>_V \,.
\end{align}
\citet{Croker2019} argue that this solves the famous averaging problem in cosmology (see for example \citet{Green2014, Buchert2015}).
Besides this, Eq.~\eqref{eq:friedmann} has important phenomenological consequences.

Namely, Eq.~\eqref{eq:friedmann} implies that the interiors of all matter objects contribute to the cosmologically effective equation of state $w_{\mathrm{eff}} \equiv P_{\mathrm{eff}} / \rho_{\mathrm{eff}}$.
Now suppose that black holes are actually gravastars or other exotic objects with an equation of state $w = -1$ in their interior.
Then, the contributions from black holes in the spatial averages Eq.~\eqref{eq:avgs} push the cosmologically effective equation of state $w_{\mathrm{eff}}$ towards $w_{\mathrm{eff}} = -1$.

Suppose further that these $w=-1$ black hole interiors are the dominant contribution in the averages Eq.~\eqref{eq:avgs}.
Then, black holes may be responsible for the accelerated expansion of the universe \citep{Croker2019}.

An implication of this scenario for dark energy (besides cosmic acceleration) is that black hole masses grow as $a^3$.
This follows from conservation of energy and is the reason why the observational results about black hole  growth from \citet{Farrah2023} may be seen as evidence for a connection between black holes and dark energy.

\section{Where it goes wrong}

Unfortunately, the first step in the argument outlined above does not work.
As I will explain, there is a mistake in the derivation of the Friedmann-like equation Eq.~\eqref{eq:friedmann} from the Einstein-Hilbert action (for other critiques see \citet{Parnovsky2023, Avelino2023, Wang2023}).

\citet{Croker2019} are interested in metrics of the form
\begin{align}
\label{eq:ansatz}
g_{\mu \nu}(\eta, \vec{x}) = a^2(\eta) \left[\eta_{\mu \nu} + \mathcal{O}(\epsilon) \right] \,,
\end{align}
where $\eta_{\mu \nu}$ is the Minkowski metric and $\epsilon$ parametrizes the inhomogeneous parts of the metric.
They plug this ansatz into the Einstein-Hilbert action and vary the action to find the equation of motion.
To lowest order in $\epsilon$, they find for the variation of the action,
\begin{align}
\label{eq:dS}
\delta S = \int d\eta \int_V d^3 \vec{x} \left[
  -\frac{3}{4 \pi G} (\delta a) \, \partial^\mu \partial_\mu a
  + a^3 T^\mu_\mu(\eta, \vec{x}) \delta a
  \right] \,,
\end{align}
where $\delta a$ is the variation of $a$.

Usually, to find the equation of motion for $a$, one would apply the principle of least action.
That is, one would require that $\delta S = 0$ for arbitrary $\delta a(\eta, \vec{x})$ and find
\begin{align}
 \label{eq:friedmannnorestriction}
 \partial_\mu \partial^\mu a = \frac{4 \pi G}{3} a^3 T^\mu_\mu \,.
\end{align}
In general, this is not consistent with the ansatz Eq.~\eqref{eq:ansatz} which requires that $a$ depends only on $\eta$ while $T^\mu_\mu$ depends also on $\vec{x}$.
So usually one would conclude that the ansatz is wrong or that the inhomogeneous parts of $T^\mu_\mu$ should be counted to be of order $\epsilon$.\footnote{
  Or that one should have taken into account other metric components besides the conformal factor when varying the action, setting them to zero only after having derived the equations of motion.
  Or not setting them to zero at all if it is inconsistent to do so.
  In any case, my main concern here is a different one.
}

But this is not what \citet{Croker2019} do.
They argue that, since in the end one is interested in the homogeneous part $a(\eta)$ of the full metric, one should restrict both $\delta a$ and $a$ to be homogeneous already before deriving the equation of motion.
Thus, in the varied action Eq.~\eqref{eq:dS} -- before deriving the equation of motion -- they assume
\begin{align}
 \label{eq:restrict}
 \delta a = \delta a (\eta) \quad (\mathrm{i.e.\;not}\;\delta a = \delta a (\eta, \vec{x}))\,,
\end{align}
and $a = a(\eta)$.
This assumption allows to pull $\delta a$ and $a$ outside the spatial integrals in Eq.~\eqref{eq:dS},
\begin{align}
\label{eq:dS2}
\delta S = \int d\eta \, \delta a \left[
  -\frac{3}{4 \pi G} \, V \, \partial_\eta^2 a
  + a^3 \, V \, \Bigl< T^\mu_\mu(\eta, \vec{x}) \Bigr>_V
  \right] \,.
\end{align}
The principle of least action then gives the Friedmann-like equation Eq.~\eqref{eq:friedmann}.

The problem is that restricting $\delta a$ and $a$ like this before deriving the equation of motion is not allowed.
If one wants to find the extremum of the action -- as the principle of least action says one should -- one must allow arbitrary variations.
Artificially restricting the variations gives \emph{some} equations, but these are not the Euler-Lagrange equations and they do not give the extremum of the action.
I give an example that demonstrates this explicitly in Appendix~\ref{sec:toymodel}.

Of course, in some cases it \emph{is} allowed to restrict the variations as in Eq.~\eqref{eq:restrict}.
For example, if all fields and sources are homogeneous.
But note that in these cases it does not matter whether or not one restricts the variations.
One ends up with the same equation of motion either way.
This is clearly not the case here.

\citet{Croker2022} argue that another case where one is allowed to restrict the variations $\delta a$ is when there is a constraint on the solution $a$.
They further claim that cosmology is such a case, with the constraint being that the metric is homogeneous to lowest order in $\epsilon$ even if $T_{\mu \nu}$ is inhomogeneous already at this order.
I dispute this claim -- there is no such constraint in the real world.

Speaking quantum-mechanically, there is nothing that forces the paths that contribute to the path integral to be homogeneous.
That is not to say that one should never restrict these paths.
A well-known example is the double slit experiment:
In many theoretical treatments, the slits are not modeled explicitly in the action.
Instead, one simply restricts the paths to always go through the (otherwise unmodeled) slits.
But note that here there is a clear physical reason for restricting the paths, namely the unmodeled slits.
There are no unmodeled slits in cosmology, so all paths must be allowed.
Equivalently, when deriving the classical equations of motion, the variations must be unrestricted.

To sum up, it is entirely possible that General Relativity produces metrics that look approximately homogeneous on large scales despite $T_{\mu \nu}$ being inhomogeneous.
But if so, that should be a consequence of extremizing the Einstein-Hilbert action in the standard way, i.e. without any restrictions on the metric variations.
It should not be artificially enforced by restricting these variations, i.e. by modifying the principle of least action.
Thus, the averaging problem remains open and black holes cannot be the dark energy, at least not in the way envisioned in \citet{Croker2019}.

\section*{Acknowledgements}
\label{sec:acknowledgements}

I thank Sabine Hossenfelder for helpful discussions and Jim Cline for pointing out a typo.
Funded by the Deutsche Forschungsgemeinschaft (DFG, German Research Foundation) – 514562826.

\begin{appendices}

\section{Toy example}
\label{sec:toymodel}

Consider the action
\begin{align}
S = \int d\eta \int_V d^3\vec{x} \left(\frac14 \phi^4(\eta, \vec{x}) - \phi(\eta, \vec{x}) T(\eta, \vec{x}) \right) \,,
\end{align}
where $\phi$ is a field and $T$ is an external source.
The variation of the action reads
\begin{align}
 \delta S = \int d\eta \int_V d^3\vec{x} \, \delta \phi \left(\phi^3 - T\right) \,.
\end{align}
I assume for simplicity that $T$ is positive.
If, as usual, one does not put any restrictions on the variation $\delta \phi$, the principle of least action gives the exact solution
\begin{align}
\phi(\eta, \vec{x}) = T(\eta, \vec{x})^{1/3} \equiv \phi_{\mathrm{exact}}(\eta, \vec{x}) \,.
\end{align}
If, instead, one uses the ansatz $\phi(\eta) (1 + \mathcal{O}(\epsilon))$ and restricts the variations $\delta \phi$ to be homogeneous, ond finds to lowest order in $\epsilon$
\begin{align}
 \phi(\eta) = \left(\frac{1}{V} \int_V d^3 \vec{x} \, T(\eta, \vec{x})\right)^{1/3} + \mathcal{O}(\epsilon) \equiv \phi_{\mathrm{restricted}}(\eta) \,.
\end{align}
The exact solution $\phi_{\mathrm{exact}}$ -- the extremum of the action -- does not agree with the solution $\phi_{\mathrm{restricted}}$ obtained by restricting the variations, if $T$ is inhomogeneous already at $\mathcal{O}(\epsilon^0)$.

\end{appendices}

\appto{\bibsetup}{\sloppy}

\printbibliography[heading=bibintoc]

\end{document}